\documentclass[twocolumn,nofootinbib, showkeys, prd,superscriptaddress, 10 pt]{revtex4}
\usepackage{amsmath}
\usepackage{amsfonts}
\usepackage{amssymb}
\usepackage{graphicx}
\usepackage{epstopdf}
\usepackage{subfigure}
\usepackage{placeins}
\begin{document}

\title{A trace of inflation in the local behavior of cosmological constant}

\author{E. Benedetto}
\email{elmobenedetto@libero.it}
\affiliation{Department of Engineering, University of Sannio, Piazza
Roma 21, 82100--Benevento, Italy.}

\author{A. Feoli}
\email{feoli@unisannio.it}
\affiliation{Department of Engineering, University of Sannio, Piazza
Roma 21, 82100--Benevento, Italy.}

\author{Liberato Pizza}
\affiliation{Dipartimento di Fisica, Universit\`a di Pisa, Largo Bruno Pontecorvo, i-56127, Pisa, Italy.}
\affiliation{Istituto Nazionale di Fisica Nucleare (INFN), Sezione di Pisa, Largo Bruno Pontecorvo, i-56127, Pisa, Italy.}
\email{liberato.pizza@df.unipi.it}


\begin{abstract}
Assuming the existence of a cosmological
constant depending on time, we study the evolution of this field in a local
region of spacetime. Solving the standard equations of Einstein Relativity
in the weak field approximation we find two asymptotes in the behavior of
the cosmological constant. Their meaning is the existence of an inflationary
era both in the far past and in the future. A trace of the initial
acceleration of the Universe can be found also in the local behavior of
cosmological constant.

\end{abstract}


\maketitle

\section{Introduction}
The recent discovery of the accelerated expansion of the Universe, obtained
studying the redshift of Supernovae \citep{Perlmutter,Riess} and the Cosmic
Microwave Background Radiation \citep{Planck}, can be understood introducing
a cosmological constant $\Lambda $ in the standard Friedman model. The
resulting $\Lambda CDM$ model is so used to describe the evolution of the
Universe with three fundamental ingredients: Dark Energy due to the
cosmological constant (69.2\%), Cold Dark Matter and Baryonic Matter
(30.8\%). In the standard $\Lambda CDM$ model, the cosmological constant
does not depend on time, but the model must admit a period of inflation at
the very beginning that cannot be explained with a very small and constant $%
\Lambda $. Immediately after the big bang, scalar fields or dilaton fields
or "ad hoc" large values of $\Lambda $ were introduced to obtain the
inflation era. Our aim is to see what we can learn about the cosmological
constant studying its local behavior. To this aim, we assume that the
cosmological constant depends on time, as  several authors have already done before
us (see for example: \citet{Overduin, ciaociao}), and that its value today is $\Lambda
_{0}\simeq 1.1\times 10^{-52}m^{-2}$ (as estimated by Planck satellite data
of the Cosmic Background Radiation \citep{Planck}), even if it may have been
different in a far past. Anyway, if the value of the cosmological constant
is small, the effects caused by its existence on the local geometry can be
treated in the weak field approximation \citep{Bernabeu}.

\section{The model}

Using the notations of Landau and Lifsits \citep{Landau}, we start from the
Einstein field equations in the form
\begin{equation}
R_{\mu \nu }=(T_{\mu \nu })_{eff}-\frac{1}{2}g_{\mu \nu }(T)_{eff},
\end{equation}%
and from a metric
\begin{equation}
g_{\mu \nu }=\eta _{\mu \nu }+h_{\mu \nu },
\end{equation}%
where $\eta _{\mu \nu }=(1,-1,-1,-1)$ is the flat Minkowski metric and we
obtain the linearized Einstein field equations neglecting terms $|h_{\mu \nu
}|^{2}<<1$
\begin{equation}
\square h_{\mu \nu }=-2(T_{\mu \nu })_{eff}+g_{\mu \nu }(T)_{eff},
\end{equation}%
where $\square =\partial _{\mu }\partial ^{\mu }$ and
\begin{equation}
(T_{\mu \nu })_{eff}=\frac{8\pi G}{c^{4}}T_{\mu \nu }+\Lambda (t)g_{\mu \nu
}.
\end{equation}%
The equation (3) was obtained imposing the gauge condition $\partial _{\mu
}\sigma _{\nu }^{\mu }=0$ \citep{Raych,Ohanian}, where
\begin{equation}
\sigma _{\mu \nu }=h_{\mu \nu }-\frac{1}{2}\eta _{\mu \nu }h.
\end{equation}%
We focus our attention on a finite region of space of volume $V$ and we
assume a perfect fluid energy - momentum tensor in the form
\begin{equation}
T_{\mu \nu }=[\rho (r)+\delta \rho (t)]u_{\mu }u_{\nu },
\end{equation}%
where there are a central source of radius $R$ with such an energy density $%
\rho (r)$ that $\rho (r)=0$ for $r>R$ and also an amount of energy density $%
\delta \rho (t)$ spread all over $V$. Furthermore $u^{\mu }\simeq (1,0,0,0)$
in the weak field approximation. Hence, being $M=\int \rho (r)d^{3}x$, the
Einstein equation (3), in the region $V$ outside the source $M$ (that is for
$r>R$), for the $00$ component can be written:
\begin{equation}
\frac{1}{c^{2}}\frac{\partial ^{2}h_{00}}{\partial t^{2}}-\nabla ^{2}h_{00}=-%
\frac{8\pi G}{c^{4}}\delta \rho (t)+2\Lambda (t)g_{00}+\frac{8\pi G}{c^{4}}%
\delta \rho (t)h_{00}.
\end{equation}%
We choose
\begin{equation}
g_{00}=1-\frac{2GM}{c^{2}r}+\alpha (t)r+\beta (t)r^{2}+\gamma (t)=1+h_{00},
\end{equation}%
where $\alpha (t)$, $\beta (t)$ and $\gamma (t)$ are unknown functions of
time. Since
\begin{equation}
\nabla ^{2}h_{00}=\frac{d^{2}h_{00}}{dr^{2}}+\frac{2}{r}\frac{dh_{00}}{dr},
\end{equation}%
from (8) we calculate
\begin{equation}
\frac{dh_{00}}{dr}=\frac{2GM}{c^{2}r^{2}}+\alpha +2\beta r,
\end{equation}%
and
\begin{equation}
\frac{d^{2}h_{00}}{dr^{2}}=-\frac{4GM}{c^{2}r^{3}}+2\beta .
\end{equation}%
Note that from the constraint $\partial _{\mu }(T_{\nu }^{\mu })_{eff}=0$ we
can add the equation
\begin{equation}
\frac{8\pi G}{c^{4}}\delta \overset{.}{\rho }+\overset{.}{\Lambda }=0,
\end{equation}%
where a dot means derivative with respect to $ct$, and we have
\begin{equation}
\frac{8\pi G}{c^{4}}\delta \rho =A-\Lambda (t),
\end{equation}%
where the constant $A$ is such that $\delta \rho >0$. Then putting $\tilde{%
\Lambda}(t)=A+\Lambda (t)$ and substituting (8), (9), (10), (11) and (13) in
the equation (7), we obtain
\begin{multline}
\overset{..}{\alpha }r+\overset{..}{\beta }r^{2}+\overset{..}{\gamma }%
-6\beta -\frac{2\alpha }{r}=\\-4A+3\tilde{\Lambda}-\frac{2GM\tilde{\Lambda}}{%
c^{2}r}+\tilde{\Lambda}\alpha r+\tilde{\Lambda}\beta r^{2}+\tilde{\Lambda}%
\gamma .
\end{multline}%
Finally we can find the unknown $\alpha (t)$, $\beta (t)$ and $\tilde{\Lambda%
}(t)$ functions from the conditions:
\begin{equation}
\left\{
\begin{array}{c}
-6\beta =3\tilde{\Lambda} \\
-2\alpha =-\frac{2GM\tilde{\Lambda}}{c^{2}} \\
\overset{..}{\alpha }=\tilde{\Lambda}\alpha \\
\overset{..}{\beta }=\tilde{\Lambda}\beta%
\end{array}%
\right. \Rightarrow \left\{
\begin{array}{c}
\beta =-\frac{\tilde{\Lambda}}{2} \\
\alpha =\frac{GM\tilde{\Lambda}}{c^{2}} \\
\frac{GM\overset{..}{\tilde{\Lambda}}}{c^{2}}=\frac{GM\tilde{\Lambda}^{2}}{%
c^{2}} \\
-\frac{\overset{..}{\tilde{\Lambda}}}{2}=-\frac{\tilde{\Lambda}^{2}}{2}%
\end{array}%
\right. .
\end{equation}%
Solving the differential equation, i.e.,
\begin{equation}
\overset{..}{\tilde{\Lambda}}=\tilde{\Lambda}^{2}\Rightarrow \tilde{\Lambda}%
(t)=6^{1/3}\wp \left[ \frac{ct+C_{1}}{6^{1/3}},(0,C_{2})\right] ,
\end{equation}%
we have derived the behavior of the cosmological constant as a function of
time in terms of the Weierstrass $\wp $ function and two integration
constants: $C_{1}$ and $C_{2}$. A plot of the function $\Lambda =\tilde{%
\Lambda}-A$ is shown in Figure 1.

More difficult is to find $\gamma (t)$ from the remaining equation:
\begin{equation}
\overset{..}{\gamma }=-4A+\tilde{\Lambda}\gamma ,
\end{equation}%
that can be solved by a numerical integration. A particular solution of this
differential equation, in the simple case of $C_{2}=0$ (when $\tilde{\Lambda}%
=6/c^{2}t^{2}$), is
\begin{equation}
\gamma =\frac{6A}{\tilde{\Lambda}}-\frac{N\tilde{\Lambda}}{A}=\frac{6}{%
1+\Lambda /A}-N(1+\Lambda /A)
\end{equation}%
and, considering that in our model $A>\Lambda $, with a suitable choice of
the constant $N$, we obtain $|\gamma (t)|<1$ in a large range of time and
the linear approximation works.

\begin{figure}[tbp]
\includegraphics[scale=0.95]{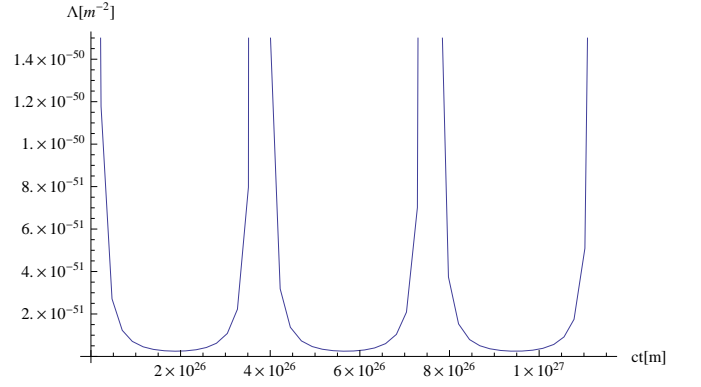}
\caption{The Weierstrass elliptic $\wp $ function in the case of $C_{1}=0$, $%
C_{2}=10^{-155}$ and $A=0$. We highlight the periodic behavior of such
function. }
\end{figure}
\begin{figure}[tbp]
\includegraphics[scale=0.95]{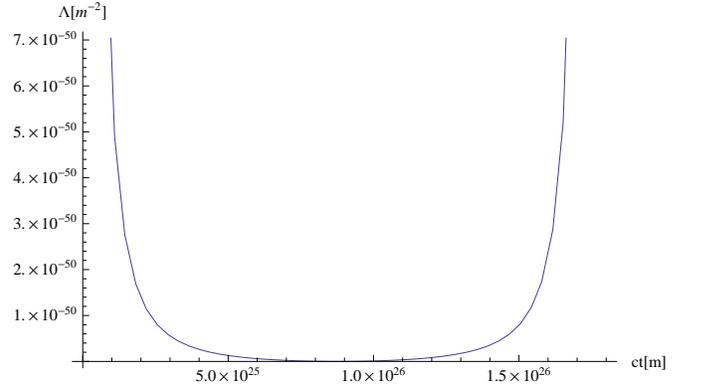}
\caption{An example of a possible solution for the cosmological constant $%
\Lambda (t)$ fixing the constants as $C_{1}=0$, $C_{2}=10^{-153}$ and $%
A\approx 10^{-51}m^{-2}$. We highlight how, for these values of $C_{1}$, $%
C_{2}$ and $A$, the function at today time ($t=13.7$ billions of years,
i.e., $ct\approx 1.3\times 10^{26}m$) acquires the expected value of $%
\Lambda _{0}=10^{-52}m^{-2}$. This value is in proximity of the minimum of
the function.}
\end{figure}
\begin{figure}[tbp]
\includegraphics[scale=0.95]{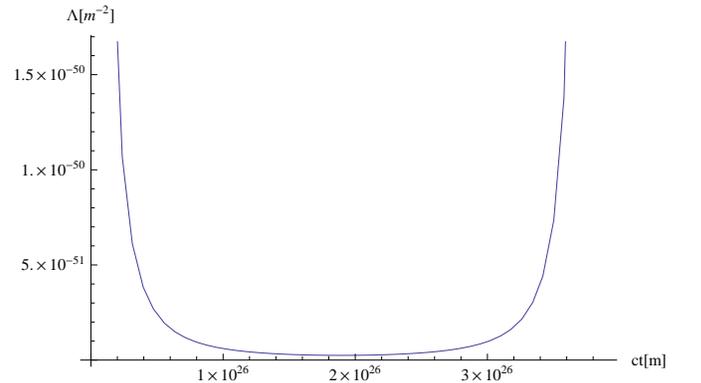}
\caption{The cosmological constant $\Lambda (t)$ obtained in the limit case $%
A=0$, fixing the constants as $C_{1}=0$, $C_{2}=10^{-155}$. We highlight
how, for these values of $C_{1}$, $C_{2}$, the cosmological constant at
today time ($t=13.7$ billions of years, i.e. $ct\approx 1.3\times 10^{26}m$)
acquires the expected value of $\Lambda _{0}=10^{-52}m^{-2}$. This value is
in proximity of the minimum of the function. }
\end{figure}

\section{Conclusions}

We obtained a first result determining the $g_{00}$ component of the metric
tensor (8) that is:
\begin{equation}
g_{00}=1-\frac{2GM}{c^{2}r}+\frac{GM\tilde{\Lambda}(t)}{c^{2}}r-\frac{\tilde{%
\Lambda}(t)}{2}r^{2}+\gamma (t),
\end{equation}%
where the fourth term induces locally a repulsive effect, the same that
globally produces the accelerating expansion of the Universe. If we compare
this term with the corresponding term in the Schwarzschild - de Sitter
spacetime \citep{Klein} we interpret the coefficient before $r^{2}$ as an
effective cosmological constant $\Lambda _{eff}=3\tilde{\Lambda}/2$.

But the plot (Figure 1) of Weierstrass elliptic $\wp$ - function is even more
interesting. It shows a periodic function that is almost constant between
two asymptotes. If the first asymptote is in $t \simeq 0$ (it occurs fixing $%
C_1 \simeq 0$), the rapid decrease of the cosmological constant seems to
denote the post inflation phase of the Universe from which, a long era of
almost constant value of $\Lambda$ starts. The amplitude of this era is
ruled by the constant $C_2$ that can be tuned to obtain a lapse of time at
least equal to the age of the universe. Finally, the constant $A$ can be
chosen in such a way that the minimum of $\Lambda(t)$ approximatively
corresponds to the today value of the cosmological constant. This way,
fixing in a suitable manner the arbitrary constants $A$, $C_1$ and $C_2$, it
is always possible to make the model in agreement with the experimental
cosmological constraints: an example is shown in Figure 2. But it is
surprising that the model works also in the case of vanishing $A$ and $%
\gamma $. We show in Figure 3 the plot of the cosmological constant obtained
fixing only $C_1$ and $C_2$ and the cosmological constraints are satisfied
as well. This case is mathematically more appealing because the number of
free adjustable parameters is smaller, but it must admit the existence of
negative energy density $\delta\rho<0$ that, anyway, is often invoked in
several physical theories, such as Casimir effect, traversable wormholes,
warp drives, black hole evaporation, or as a property of some kind of exotic
matter.

Independently on the choice of the constants, the $\Lambda $ function has
always the two asymptotes (except the trivial case $C_{2}=0$ with only one
asymptote for $t\rightarrow 0$) whose meaning is an inflationary behavior of
the cosmological constant. Of course, going back into the past, when the
cosmological constant becomes larger and larger, there will be an instant
when the linear approximation can be no longer applied. Anyway we consider
an important result to have obtained, with a so simple model and without
introducing any other auxiliary fields besides $\Lambda(t)$, the prediction
of a past and also of a future inflationary era.

\acknowledgments

 This work was partially supported by research funds of the University of
Sannio.


\begin{thebibliography}{Overduin \& Cooperstock(1998)}

\bibitem[Perlmutter et al.(1999)]{Perlmutter} Perlmutter, S. et al. 1999 \apj%
\textbf{517}, 565

\bibitem[Riess et al.(1998)]{Riess} Riess, A.G. et al. 1998 AJ \textbf{116},
1009

\bibitem[Ade et al.(2015)]{Planck} Ade, P.A.R. et al. 2015 results. XIII
Cosmological parameters (Planck Collaboration):arXiv:1502.01589

\bibitem[Feng \& Li(2014)]{ciaociao} Feng, C.-J. \& Li, X.-Z. 2014 Phys.
Rev. \textbf{D90}, 103009

\bibitem[Overduin \& Cooperstock(1998)]{Overduin} Overduin, J.M., \&
Cooperstock, F.I. 1998 Phys. Rev. \textbf{D58}, 043506

\bibitem[Bernabeu et al.(2007)]{Bernabeu} Bernabeu, J., Espinoza,C., \&
Mavromatos, N.E. 2010 Phys.Rev. \textbf{D81}, 084002

\bibitem[Landau \& Lifshitz(1975)]{Landau} Landau, L.D., \& Lifshitz, E.M.
1975 \textit{The Classical Theory of Fields} Vol. 2 (4th
ed.)(Butterworth-Heinemann)

\bibitem[Ohanian \& Ruffini(2013)]{Ohanian} Ohanian, H.C., \& Ruffini, R.
2013 \textit{Gravitation and Spacetime} (Cambridge University Press,
Cambridge)





\bibitem[Raychauduri et al.(1992)]{Raych} Raychauduri, A.K., Banerji, S., \&
Banerjee, A. 1992 \textit{General Relativity, Astrophysics and Cosmology}
(Springer -- Verlag, New York) Chapter 3


\bibitem[Klein \& Collas(2010)]{Klein} Klein, D., \& Collas, P. 2010 Phys.
Rev. \textbf{D81}, 063518





\end{thebibliography}
\end{document}